\documentclass[aps,superscriptaddress,nofootinbib,floatfix,notitlepage]{revtex4-1}
\usepackage[utf8x]{inputenc}
\pdfoutput=1
\usepackage{graphicx}
\usepackage{hyperref}
\usepackage{bm}

\begin{document}

\title{The \textbf{UT}\textit{fit} Collaboration Average of $D$ meson
  mixing data: Spring 2012
  \vspace*{0.5cm}
}

\collaboration{\begin{figure}[h!]
  \begin{center}
  \includegraphics[width=0.13\textwidth]{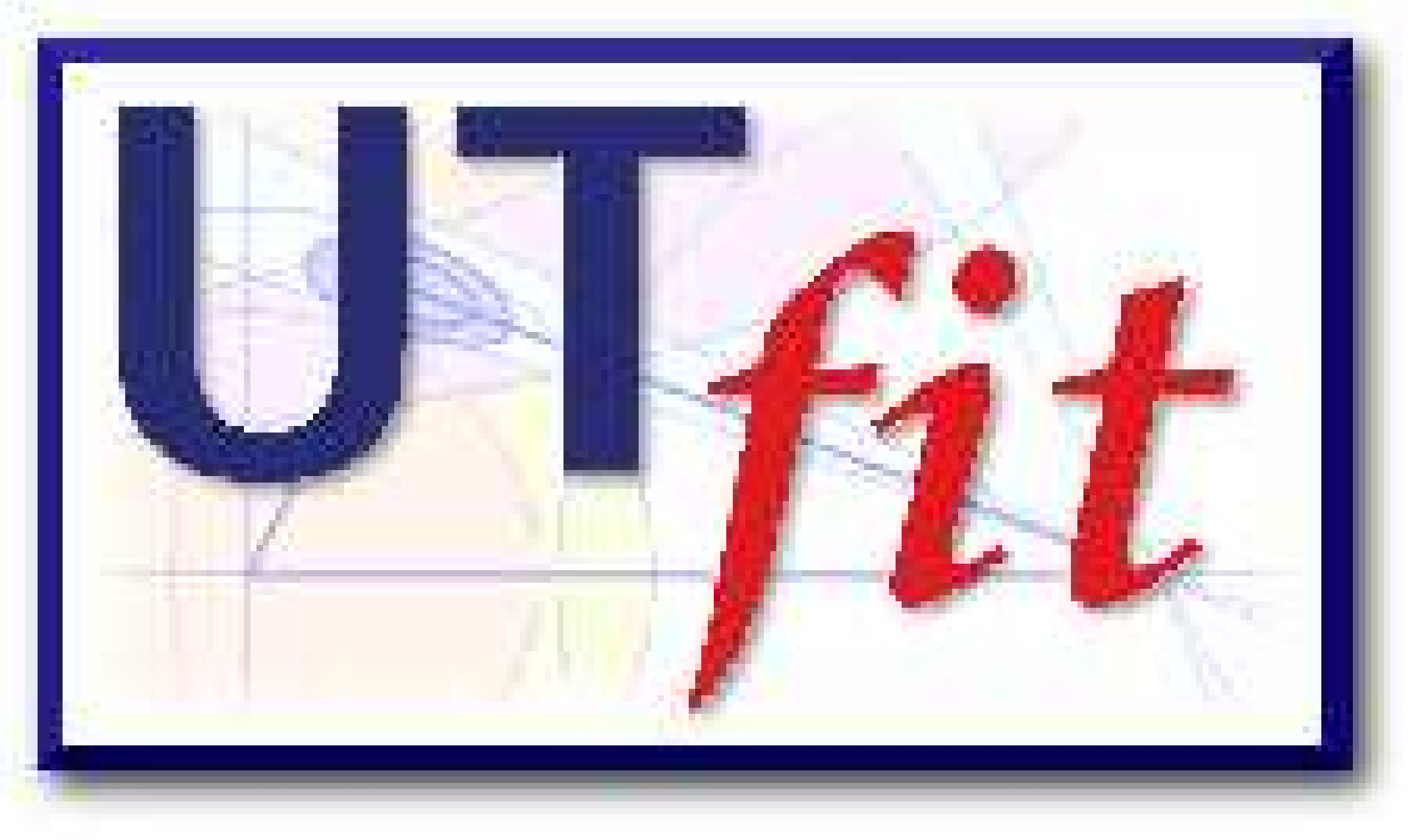}
  \end{center}
 \end{figure}\textbf{UT}\textit{fit} Collaboration
}
\homepage{http://www.utfit.org} 
\author{A.J.~Bevan}
\affiliation{Queen Mary, University of London, Mile End Road, London E1 4NS, United Kingdom}
\author{M.~Bona}
\affiliation{Queen Mary, University of London, Mile End Road, London E1 4NS, United Kingdom}
\author{M.~Ciuchini}
\affiliation{INFN,  Sezione di Roma Tre, Via della Vasca Navale 84, I-00146 Roma, Italy}
\author{D.~Derkach}
\affiliation{CERN, CH-1211 Geneva 23, Switzerland}
\author{E.~Franco}
\affiliation{INFN, Sezione di Roma, Piazzale A. Moro 2, I-00185 Roma, Italy}
\author{V.~Lubicz}
\affiliation{INFN,  Sezione di Roma Tre, Via della Vasca Navale 84, I-00146 Roma, Italy}
\affiliation{Dipartimento di Fisica, Universit{\`a} di Roma Tre, Via della Vasca Navale 84, 
  I-00146 Roma, Italy} 
\author{G.~Martinelli}
\affiliation{INFN, Sezione di Roma, Piazzale A. Moro 2, I-00185 Roma, Italy}
\affiliation{SISSA-ISAS, Via Bonomea 265, I-34136 Trieste, Italy} 
\author{F.~Parodi}
\affiliation{Dipartimento di Fisica, Universit\`a di Genova and INFN, Via Dodecaneso 33, I-16146
  Genova, Italy} 
\author{M.~Pierini}
\affiliation{CERN, CH-1211 Geneva 23, Switzerland}
\author{C.~Schiavi}
\affiliation{Dipartimento di Fisica, Universit\`a di Genova and INFN, Via Dodecaneso 33, I-16146
  Genova, Italy} 
\author{L.~Silvestrini}
\affiliation{INFN, Sezione di Roma, Piazzale A. Moro 2, I-00185 Roma, Italy}
\author{V.~Sordini}
\affiliation{IPNL-IN2P3, 4 Rue Enrico Fermi, F-69622 Villeurbanne Cedex, France}
\author{A.~Stocchi}
\affiliation{Laboratoire de l'Acc\'el\'erateur Lin\'eaire, IN2P3-CNRS et
  Universit\'e de Paris-Sud, BP 34, F-91898 Orsay Cedex, France}
\author{C.~Tarantino}
\affiliation{INFN,  Sezione di Roma Tre, Via della Vasca Navale 84, I-00146 Roma, Italy}
\affiliation{Dipartimento di Fisica, Universit{\`a} di Roma Tre, Via della Vasca Navale 84,
  I-00146 Roma, Italy} 
\author{V.~Vagnoni}
\affiliation{INFN, Sezione di Bologna,  Via Irnerio 46, I-40126 Bologna, Italy}

\begin{abstract}
  We derive constraints on the parameters $M_{12}$, $\Gamma_{12}$ and
  $\Phi_{12}$ that describe $D$ meson mixing using all available data,
  allowing for CP violation. We also provide posterior distributions
  and predictions for observable parameters appearing in $D$ physics.
\end{abstract}
 
\maketitle

Meson-antimeson mixing in the neutral $D$ system has been established
only in 2007~\cite{hep-ex/0703020,hep-ex/0703036,0704.1000}. Early
combinations of available data allowed to put stringent constraints on
New Physics (NP) contributions, although the possibility of
non-standard CP violation remained
open~\cite{hep-ph/0703204,Nir:2007ac,Golowich:2007ka,Fajfer:2007dy,Bona:2007vi}. More
recently, CP violation in the $D$ system received considerable
attention after the measurement at hadron colliders of large direct CP
violation in $D \to \pi\pi$ and $D\to KK$ decays
\cite{Aaij:2011in,notacdf}, which may signal the presence of
NP~\cite{Brod:2011re,Pirtskhalava:2011va,Bhattacharya:2012ah,Cheng:2012wr,Franco:2012ck,Brod:2012ud}. It
then becomes crucial to extract updated information on the mixing
amplitude in order both to disentangle more precisely indirect and
direct CP violation in $D \to \pi\pi$ and $D\to KK$, and to obtain
up-to-date constraints on NP in $\Delta C=2$ transitions that can be
used to constrain NP contributions to $\Delta C=1$ processes in any
given model. 

\begin{table}[htb]
  \centering
  \begin{tabular}{|cccccccc|}
    \hline
    Observable & Value & \multicolumn{5}{c}{Correlation Coeff.} &
    Reference \\ \hline
    $y_{CP}$ & $(0.866 \pm 0.155)\%$ & & & & & & \cite{hep-ex/0004034,hep-ex/0111024,hep-ex/0111026,hep-ex/0703036,0712.2249,0908.0761,0905.4185,Aaij:2011ad,Staric,Neri} \\ \hline
    $A_\Gamma$ & $(0.022 \pm 0.161)\%$ & & & & & & \cite{hep-ex/9903012,hep-ex/0703036,0712.2249,Aaij:2011ad,Staric,Neri} \\ \hline
    $x$ & $(0.811 \pm 0.334)\%$ & 1 & -0.007 & -0.255$\alpha$ &
    0.216 & & \cite{0704.1000} \\
    $y$ & $(0.309 \pm 0.281)\%$ & -0.007 & 1 & -0.019$\alpha$ &
    -0.280 & & \cite{0704.1000} \\
    $\vert q/p \vert$ & $(0.95 \pm 0.22 \pm 0.10)\%$ & -0.255$\alpha$
    & -0.019$\alpha$ & 1 &
    -0.128 $\alpha$ & & \cite{0704.1000} \\
    $\phi$ & $(-0.035 \pm 0.19 \pm 0.09)$ & 0.216
    & -0.280 &
    -0.128 $\alpha$ & 1 & & \cite{0704.1000} \\\hline
    $x$ & $(0.16 \pm 0.23 \pm 0.12 \pm 0.08)\%$ & 1 & 0.0615 & & & & \cite{1004.5053} \\
    $y$ & $(0.57 \pm 0.20 \pm 0.13 \pm 0.07)\%$ & 0.0615 & 1 & & & & \cite{1004.5053} \\
    \hline
    $R_M$ & $(0.0130 \pm 0.0269)\%$ & & & & & & \cite{hep-ex/9606016,hep-ex/0502012,hep-ex/0408066,0705.0704,Bitenc:2008bk} \\\hline
    $(x^\prime_+)_{K\pi\pi^0}$ & $(2.48 \pm 0.59 \pm 0.39)\%$ & 1 & -0.69 &  &   &
    & \cite{0807.4544} \\
    $(y^\prime_+)_{K\pi\pi^0}$ & $(-0.07 \pm 0.65 \pm 0.50)\%$ & -0.69 &
    1 &  &   &
    & \cite{0807.4544} \\
    $(x^\prime_-)_{K\pi\pi^0}$ & $(3.50 \pm 0.78 \pm 0.65)\%$ & 1 & -0.66 &  &   &
    & \cite{0807.4544} \\
    $(y^\prime_-)_{K\pi\pi^0}$ & $(-0.82 \pm 0.68 \pm 0.41)\%$ & -0.66 &
    1 &  &   &
    & \cite{0807.4544} \\ \hline
    $x^2$ & $(0.1549 \pm 0.2223)\%$ & 1 & -0.6217 & -0.00224 & 0.3698
    & 0.01567
    & \cite{cleoc} \\
    $y$ & $(2.997 \pm 2.293)\%$ & -0.6217 & 1 & 0.00414 & -0.5756  & -0.0243
    & \cite{cleoc} \\ 
    $R_D$ & $(0.4118 \pm 0.0948)\%$ & -0.00224 & 0.00414 & 1 & 0.0035
    & 0.00978
    & \cite{cleoc} \\
    $2\sqrt{R_D}\cos\delta_{K\pi}$ & $(12.64 \pm 2.86)\%$ & 0.3698 &
    -0.5756 & 0.0035 & 1  & 0.0471
    & \cite{cleoc} \\
    $2\sqrt{R_D}\sin\delta_{K\pi}$ & $(-0.5242 \pm 6.426)\%$ & 0.01567
    & -0.0243 & 0.00978 & 0.0471  & 1
    & \cite{cleoc} \\
   \hline
    $R_D$ & $(0.3030 \pm 0.0189)\%$ & 1 & 0.77 & -0.87 &   &
    & \cite{hep-ex/0703020} \\
    $(x^\prime_+)^2_{K\pi}$ & $(-0.024 \pm 0.052)\%$ & 0.77 & 1 & -0.94 &   &
    & \cite{hep-ex/0703020} \\
    $(y^\prime_+)_{K\pi}$ & $(0.98 \pm 0.78)\%$ & -0.87 & -0.94 & 1  & &
    & \cite{hep-ex/0703020} \\\hline
    $A_D$ & $(-2.1 \pm 5.4)\%$ & 1 & 0.77 & -0.87 &   &
    & \cite{hep-ex/0703020} \\
    $(x^\prime_-)^2_{K\pi}$ & $(-0.020 \pm 0.050)\%$ & 0.77 & 1 & -0.94 &   &
    & \cite{hep-ex/0703020} \\
    $(y^\prime_-)_{K\pi}$ & $(0.96 \pm 0.75)\%$ & -0.87 & -0.94 & 1  & &
    & \cite{hep-ex/0703020} \\\hline
    $R_D$ & $(0.364 \pm 0.018)\%$ & 1 & 0.655 & -0.834 &   &
    & \cite{hep-ex/0601029} \\
    $(x^\prime_+)^2_{K\pi}$ & $(0.032 \pm 0.037)\%$ & 0.655 & 1 & -0.909 &   &
    & \cite{hep-ex/0601029} \\
    $(y^\prime_+)_{K\pi}$ & $(-0.12 \pm 0.58)\%$ & -0.834 & -0.909 & 1  & &
    & \cite{hep-ex/0601029} \\\hline
    $A_D$ & $(2.3 \pm 4.7)\%$ & 1 & 0.655 & -0.834 &   &
    & \cite{hep-ex/0601029} \\
    $(x^\prime_-)^2_{K\pi}$ & $(0.006 \pm 0.034)\%$ & 0.655 & 1 & -0.909 &   &
    & \cite{hep-ex/0601029} \\
    $(y^\prime_-)_{K\pi}$ & $(0.20 \pm 0.54)\%$ & -0.834 & -0.909 & 1  & &
    & \cite{hep-ex/0601029} \\\hline
    CP asymmetry & Value & \multicolumn{5}{c}{
      $\Delta\langle t\rangle/\tau_{D^0}$} & Reference \\ \hline 
    $A_\mathrm{CP}(D^0 \to K^+K^-)$ & $(-0.24\pm
      0.24)\%$ & &&& & 
    & \cite{Aubert:2007if,Staric:2008rx} \\
    $A_\mathrm{CP}(D^0 \to \pi^+\pi^-)$ & $(0.11
      \pm 0.39)\%$ & &&& & 
    & \cite{Aubert:2007if,Staric:2008rx}\\
    $\Delta A_\mathrm{CP}$ & $(-0.82 \pm 0.21
      \pm 0.11)\%$ & \multicolumn{4}{c}{$(9.83 \pm 0.22
      \pm 0.19)\%$} & 
    & \cite{Aaij:2011in} \\
    $\Delta A_\mathrm{CP}$ & $(-0.62 \pm 0.21
      \pm 0.10)\%$ & \multicolumn{4}{c}{$(26 \pm 1)\%$} & 
    & \cite{notacdf} \\ \hline
  \end{tabular}
  \caption{Experimental data used in the analysis of $D$ mixing, from
    ref.~\cite{[{}][{ and online updates at
        \url{http://www.slac.stanford.edu/xorg/hfag/}}]1010.1589}. $\alpha
    = (1 + \vert q/p \vert)^2/2$ and $\Delta A_\mathrm{CP} =
    A_\mathrm{CP}(D^0 \to K^+K^-) -A_\mathrm{CP}(D^0 \to
    \pi^+\pi^-)$. Asymmetric errors have been 
    symmetrized. We do not use measurements that do not allow for CP
    violation in mixing, except for ref.~\cite{1004.5053} (as
      shown in ref.~\cite{0704.1000}, the results for $x$ and $y$ from
      the Dalitz analysis of $D \to K_s \pi \pi$ are not sensitive to
      the assumptions about CP violation in mixing).}
  \label{tab:dmixexp}
\end{table}

In this letter, we perform a fit to the experimental data in
Table~\ref{tab:dmixexp} following the statistical method described in
ref.~\cite{hep-ph/0012308}.  We assume that all Cabibbo allowed (and
doubly Cabibbo suppressed) decay amplitudes in the phase convention
$\mathrm{CP}\vert D\rangle = \vert \bar D \rangle$ and
$\mathrm{CP}\vert f\rangle = \eta_{\mathrm{CP}}^f \vert \bar f
\rangle$ satisfy the relation $\mathcal{A}(D \to f) =
\eta_{\mathrm{CP}}^f\mathcal{A}(\bar D \to \bar f)$, which is expected
to hold in the SM (in the standard CKM phase convention) with an
accuracy much better than present experimental errors. In the same
approximation this implies $\Gamma_{12}$ real. For singly Cabibbo
suppressed decays $D^0 \to K^+K^-$ and $D^0 \to \pi^+\pi^-$ we allow
for direct CP violation to be present. We assume flat priors for $x =
\Delta m_D/\Gamma_D$, $y = \Delta \Gamma_D/(2 \Gamma_D)$ and $\vert
q/p\vert$, with $\vert D_{L,S} \rangle = p \vert D^0 \rangle \pm q
\vert \bar D^0 \rangle$ and $\vert p \vert^2+\vert q \vert^2 = 1$. We
can then express all mixing-related observables in terms of $x$, $y$
and $\vert q/p\vert$ using the following
formul{\ae}~\cite{Branco:1999fs,hep-ph/0205113,hep-ph/0703204,0907.3917,Grossman:2009mn}:
\begin{eqnarray}
  \label{eq:xyandco}
  &&\delta = \frac{1 - \vert q/p \vert^2}{1+\vert q/p
    \vert^2} \,,\quad \phi = \arg(q/p) = \arg (y+i \delta x)\,,\quad
  A_M = \frac{\vert q/p \vert^4 -1}{\vert q/p
    \vert^4+1}\,, \quad
  R_M =\frac{x^2+y^2}{2}\,, \\
  &&
  \left(
    \begin{array}{c}
      x^\prime_f \\
      y^\prime_f
    \end{array}
  \right) =
  \left(
    \begin{array}{cc}
      \cos \delta_{f} & \sin \delta_{f} \\
      -\sin \delta_{f} & \cos \delta_{f}
    \end{array}
  \right)   \left(
    \begin{array}{c}
      x \\
      y
    \end{array}
  \right)
  \,,\quad
  (x^{\prime}_\pm)_f = \left\vert
    \frac{q}{p}
  \right\vert^{\pm 1}(x^\prime_f\cos \phi \pm y^\prime_f \sin
  \phi)\,, \quad 
  (y^\prime_\pm)_f   =
  \left\vert
    \frac{q}{p}
  \right\vert^{\pm 1}(y^\prime_f\cos \phi \mp x^\prime_f \sin
  \phi)\,,\nonumber \\ 
  && y_\mathrm{CP} =
  \left(
    \left\vert
      \frac{q}{p}
    \right\vert + \left\vert
      \frac{p}{q}
    \right\vert
  \right) \frac{y}{2} \cos \phi- \left(
    \left\vert
      \frac{q}{p}
    \right\vert - \left\vert
      \frac{p}{q}
    \right\vert
  \right) \frac{x}{2}\sin \phi\,,\quad A_\Gamma =  \left(
    \left\vert
      \frac{q}{p}
    \right\vert - \left\vert
      \frac{p}{q}
    \right\vert
  \right) \frac{y}{2} \cos \phi- \left(
    \left\vert
      \frac{q}{p}
    \right\vert + \left\vert
      \frac{p}{q}
    \right\vert
  \right) \frac{x}{2}\sin \phi\,, \nonumber 
 \\
   && R_D = \frac{\Gamma(D^0 \to K^+\pi^-)+\Gamma(\bar D^0 \to
     K^-\pi^+)}{\Gamma(D^0 \to K^-\pi^+)+\Gamma(\bar D^0 \to
     K^+\pi^-)}\,, 
   \quad A_{D} = \frac{\Gamma(D^0 \to K^+\pi^-)-\Gamma(\bar D^0 \to
     K^-\pi^+)}{\Gamma(D^0 \to K^+\pi^-)+\Gamma(\bar D^0 \to
     K^-\pi^+)} \,,\nonumber
\end{eqnarray}
with $\delta_{f}$ a strong phase and $A_\mathrm{D}$ forced to
vanish in the fit.  In addition, for the CP asymmetries we have
\begin{equation}
A_{\rm CP}(f) = 
\frac{\Gamma(D^0\to f) - \Gamma(\bar{D}^0\to \bar{f})}
  {\Gamma(D^0\to f) + \Gamma(\bar{D}^0\to \bar{f})}
\approx
a_{\rm CP}^{\rm dir}(f) - A_{\Gamma} 
\int_0^\infty \!dt\, \frac{t}{\tau_{D^0}}\,D_f(t)
=
a_{\rm CP}^{\rm dir}(f) - \frac{\langle t\rangle_f}{\tau_{D^0}}\, 
A_\Gamma\,,
\end{equation}
where $D_f(t)$ is the observed distribution of proper decay
time and $\tau_{D^0}$ is the lifetime of the neutral $D$ mesons.  

For the purpose of constraining NP, it is useful
to express the fit results in terms of the $\Delta C=2$ effective
Hamiltonian matrix elements $M_{12}$ and $\Gamma_{12}$:
\begin{equation}
  \vert M_{12} \vert = \frac{1}{\tau_D } \sqrt{\frac{x^2+\delta^2
      y^2}{4(1-\delta^2)}}\,,\quad
  \vert \Gamma_{12} \vert= \frac{1}{\tau_D }\sqrt{\frac{y^2+\delta^2
      x^2}{1-\delta^2}}\,, \quad
  \sin \Phi_{12} = \frac{\vert \Gamma_{12}\vert^2 + 4 \vert
    M_{12}\vert^2 - (x^2+y^2)\vert q/p\vert^2/\tau_D^2}{4 \vert M_{12}
    \Gamma_{12}\vert}\,,
  \label{eq:m12g12}
\end{equation}
with $\Phi_{12}=\arg \Gamma_{12}/M_{12}$. Consistently with the
assumption $\mathcal{A}(D \to f) = \mathcal{A}(\bar D \to \bar f)$,
$\Gamma_{12}$ can be taken real with negligible NP contributions, and
a nonvanishing $\Phi_{12}$ can be interpreted as a signal of new
sources of CP violation in $M_{12}$. For the sake of completeness, we
report here also the formul{\ae} to compute the observables $x$, $y$ and
$\delta$ from $M_{12}$ and $\Gamma_{12}$:
\begin{eqnarray}
  \sqrt{2}\, \Delta m &=& \mathrm{sign}(\cos\Phi_{12}) \sqrt{4 \vert
    M_{12} \vert^2 - \vert 
    \Gamma_{12} \vert^2 + \sqrt{(4\vert M_{12} \vert^2+ \vert
      \Gamma_{12} \vert^2)^2-16 \vert M_{12} \vert^2 \vert
      \Gamma_{12} \vert^2 \sin^2\Phi_{12}} }\,,\nonumber \\
  \sqrt{2}\, \Delta \Gamma&=& \sqrt{\vert \Gamma_{12} \vert^2 - 4\vert
    M_{12} \vert^2 + \sqrt{(4\vert M_{12} \vert^2+ \vert
      \Gamma_{12} \vert^2)^2-16 \vert M_{12} \vert^2 \vert
      \Gamma_{12} \vert^2 \sin^2\Phi_{12}} }\,,\nonumber \\
  \delta &=&\frac{ 2 \vert M_{12} \vert \vert
    \Gamma_{12} \vert \sin\Phi_{12}}{(\Delta m)^2 + \vert
    \Gamma_{12}\vert^2} \,,
  \label{eq:m12g12inv}
\end{eqnarray}
in agreement with \cite{0907.3917} up to a factor of $\sqrt{2}$. 

\begin{table}[t]
  \centering
  \begin{tabular}{|ccc|}
    \hline
    parameter & result @ $68\%$ prob. & $95\%$ prob. range \\
    \hline
    $\vert M_{12}\vert$ [1/ps] & $(6.9 \pm 2.4) \cdot 10^{-3}$ & $[2.1,11.5]
    \cdot 10^{-3}$ \\
    $\vert \Gamma_{12}\vert$ [1/ps] & $(17.2 \pm 2.5) \cdot 10^{-3}$ &
    $[12.3,22.4] 
    \cdot 10^{-3}$ \\
    $\Phi_{12}$ [$^\circ$] & $(-6 \pm 9)$ &
    $[-37,13]$ \\
    \hline
    $x$ & $(5.6 \pm 2.0) \cdot 10^{-3}$ & $[1.4,9.6] \cdot 10^{-3}$ \\
    $y$ & $(7.0 \pm 1.0) \cdot 10^{-3}$ & $[5.0,9.1] \cdot 10^{-3}$ \\
    $\vert q/p \vert -1$ & $(5.3 \pm 7.7) \cdot 10^{-2}$ & $[-8.5,25.6] \cdot 10^{-2}$ \\
    $\phi$ [$^\circ$] & $(-2.4 \pm 2.9)$ &
    $[-8.8,3.7]$ \\
    \hline
    $A_\Gamma$ & $(0.7 \pm 0.8)\cdot 10^{-3} $ &
    $[-0.9,2.3]\cdot 10^{-3} $ \\ 
    $A_M$ &  $(11 \pm 14) \cdot 10^{-2} $ & $[-15,44] \cdot
    10^{-2}$ \\
    $R_M$ &  $(4.0 \pm 1.4) \cdot 10^{-5} $ & $[1.7,7.2] \cdot
    10^{-5}$ \\
    $R_D$ & $(3.27 \pm 0.08) \cdot 10^{-3} $ & $[3.10,3.44] \cdot
    10^{-3}$ \\
    $\delta_{K\pi}$ [$^\circ$] & $(18 \pm 12)$  &
    $[-14,40]$ \\
    $\delta_{K\pi\pi^0}$ [$^\circ$] & $(31 \pm 20)$  &
    $[-11,73]$ \\ \hline
    $a_\mathrm{CP}^\mathrm{dir}(D^0 \to K^+K^-)$ & $(-2.6 \pm 2.2) \cdot
    10^{-3}$ & $[-7.1,1.9]  \cdot
    10^{-3}$ \\
    $a_\mathrm{CP}^\mathrm{dir}(D^0 \to \pi^+\pi^-)$ & $(4.1 \pm 2.4) \cdot
    10^{-3}$ & $[-0.8,9.0]  \cdot
    10^{-3}$ \\
    $\Delta a_\mathrm{CP}^\mathrm{dir}$ &  $(6.6 \pm 1.6) \cdot
    10^{-3}$ & $[-9.8,3.5]  \cdot
    10^{-3}$ \\ \hline
  \end{tabular}
  \caption{Results of the fit to $D$ mixing data. $\Delta
    a_\mathrm{CP}^\mathrm{dir} = a_\mathrm{CP}^\mathrm{dir}(D^0 \to
    K^+K^-)-a_\mathrm{CP}^\mathrm{dir}(D^0 \to \pi^+\pi^-)$.}
  \label{tab:ddmix_res}
\end{table}

\begin{figure}[htb]
  \centering
  \includegraphics[width=.3\textwidth]{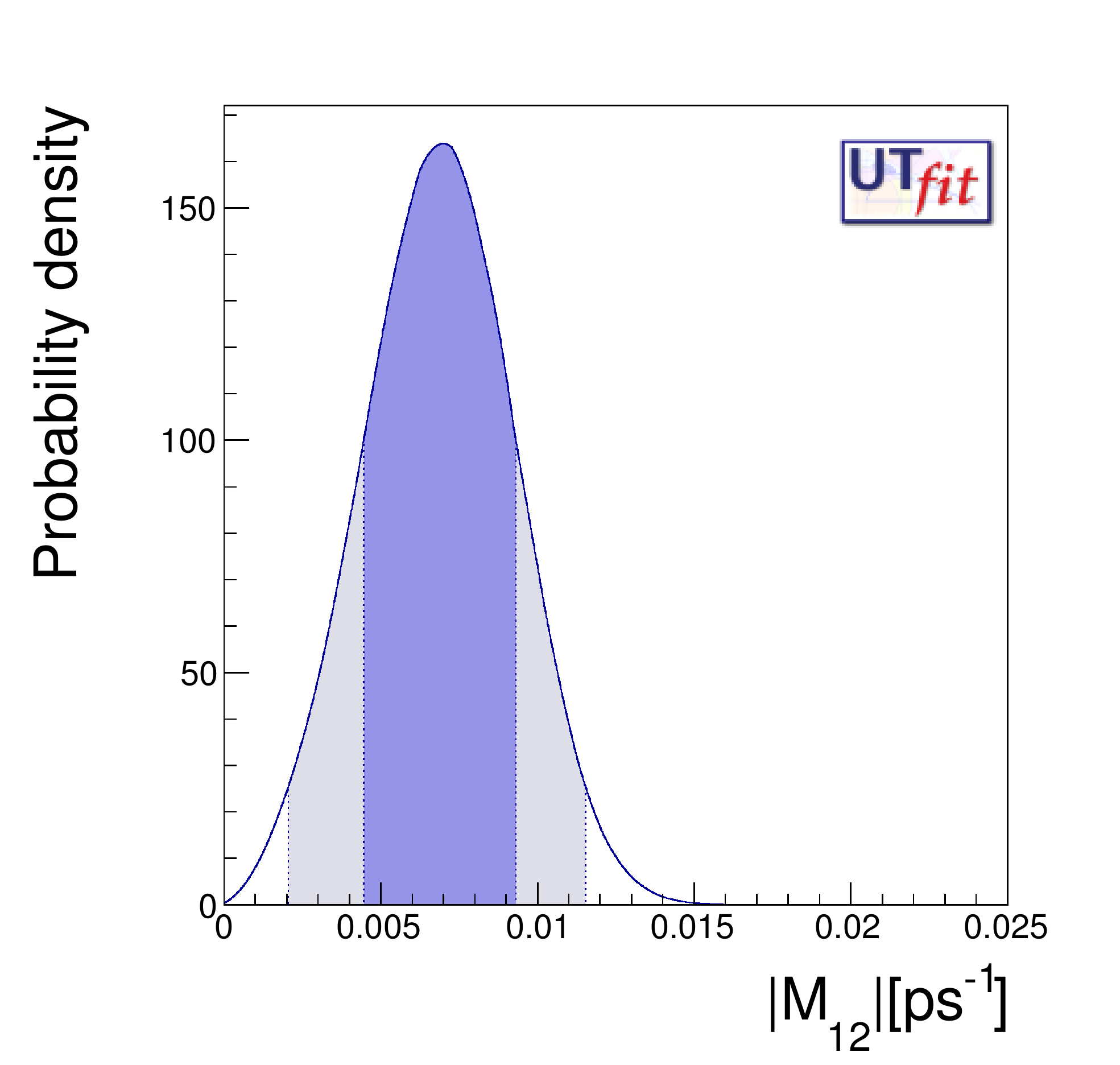}
  \includegraphics[width=.3\textwidth]{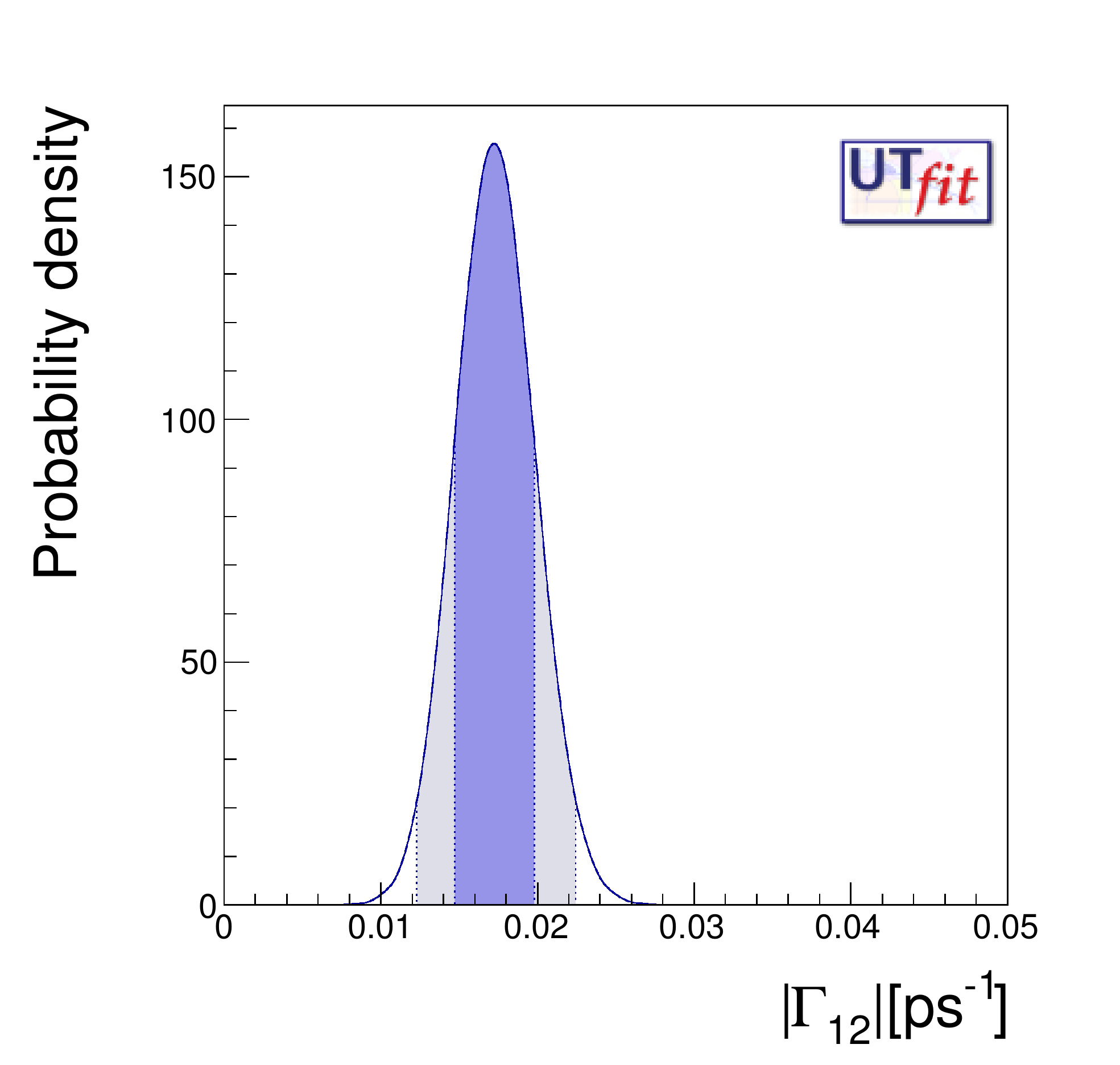}
  \includegraphics[width=.3\textwidth]{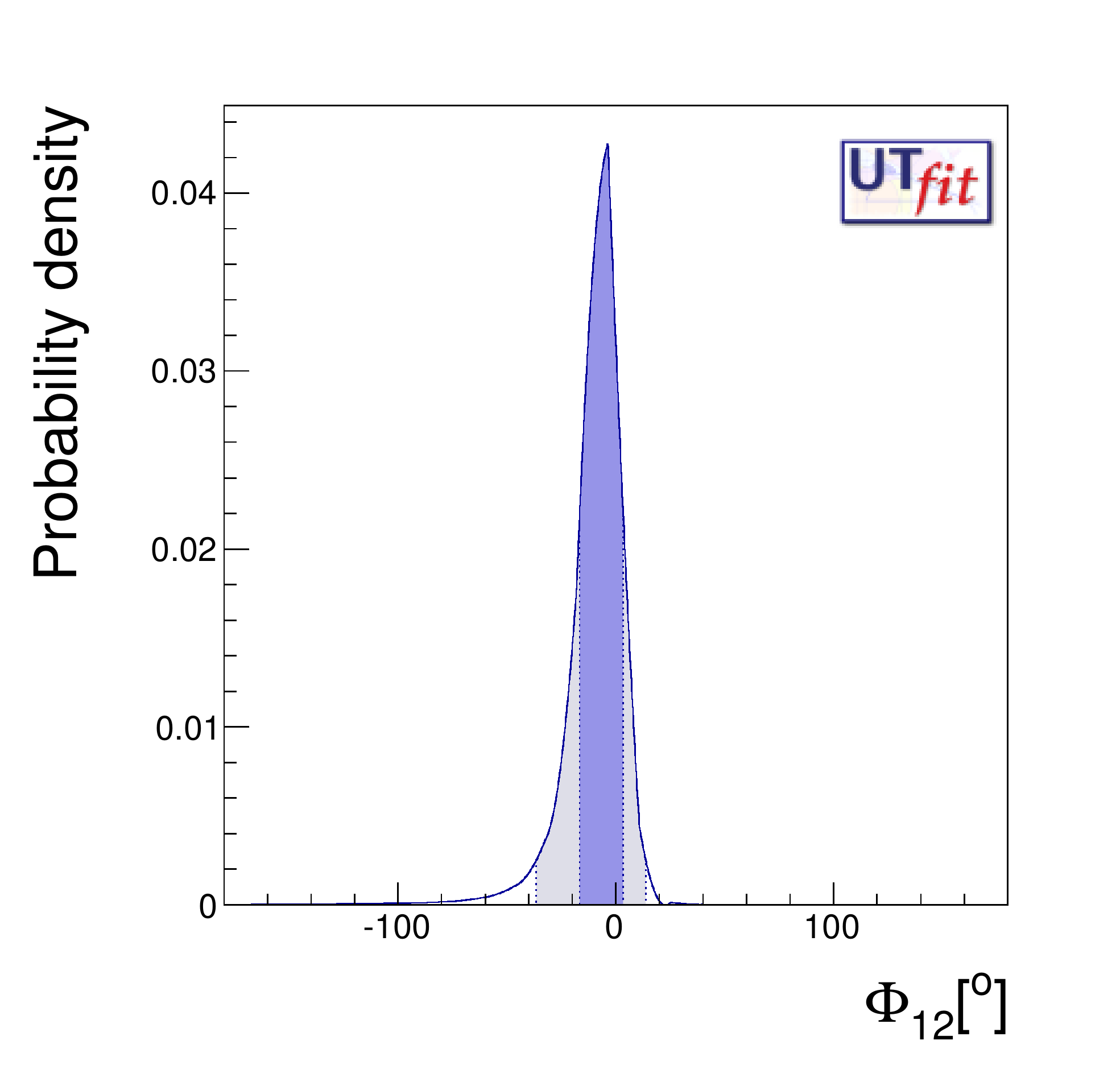}
  \caption{One-dimensional p.d.f. for the parameters $\vert M_{12}
    \vert$, $\vert \Gamma_{12}
    \vert$ and $\Phi_{12}$.}
  \label{fig:ddmix_1d}
\end{figure}

\begin{figure}[htb]
  \centering
  \includegraphics[width=.24\textwidth]{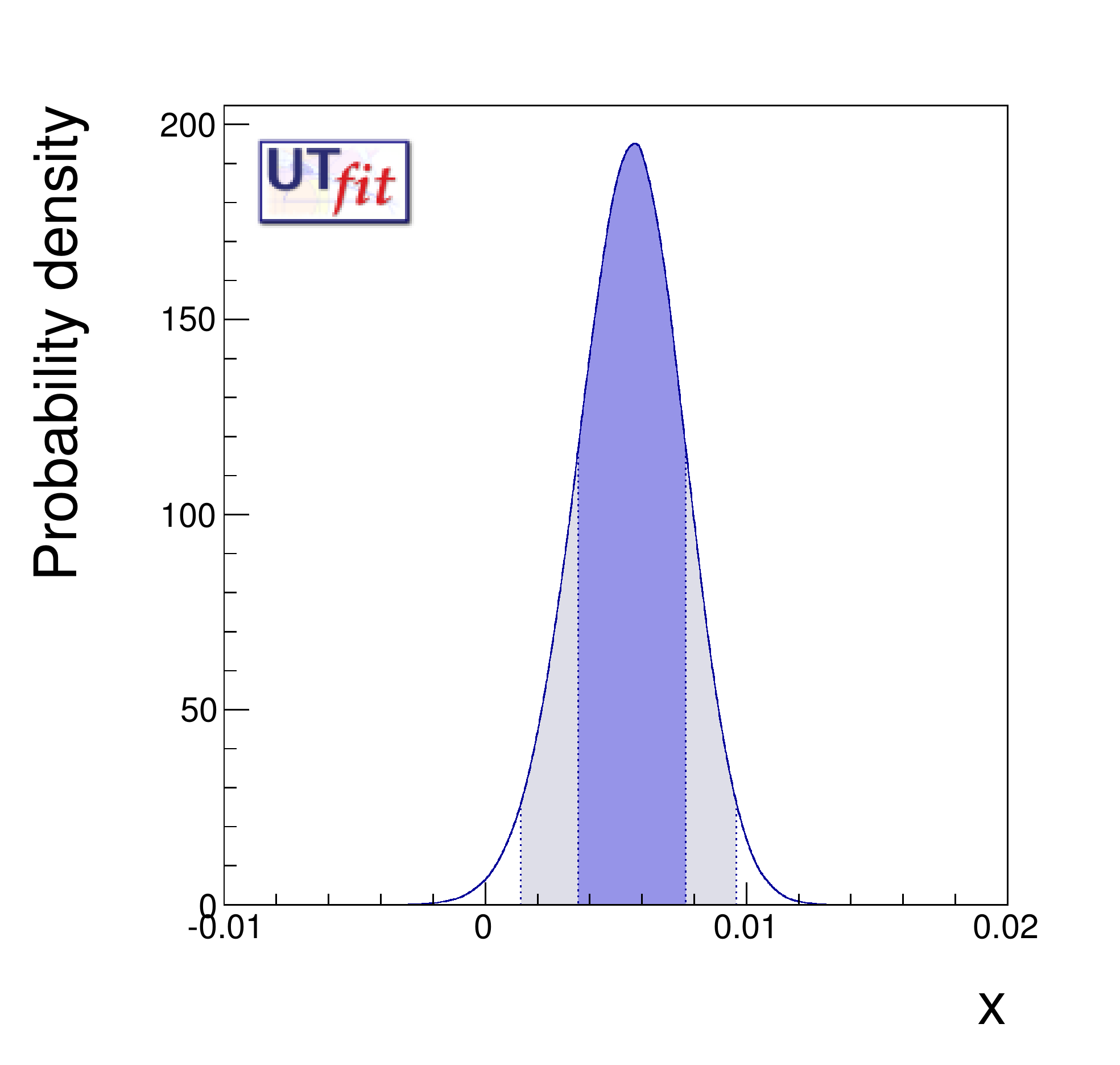}
  \includegraphics[width=.24\textwidth]{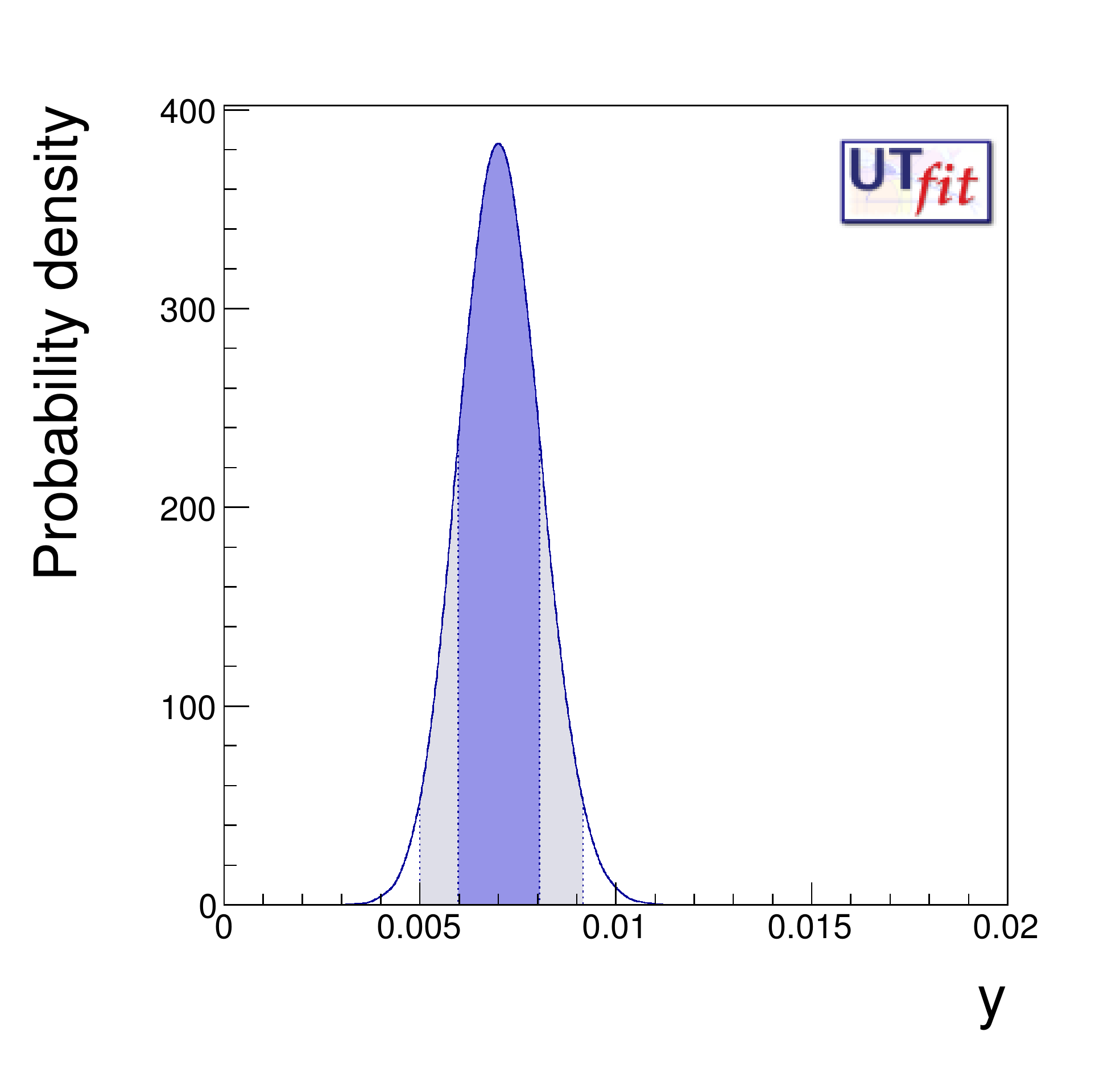}
  \includegraphics[width=.24\textwidth]{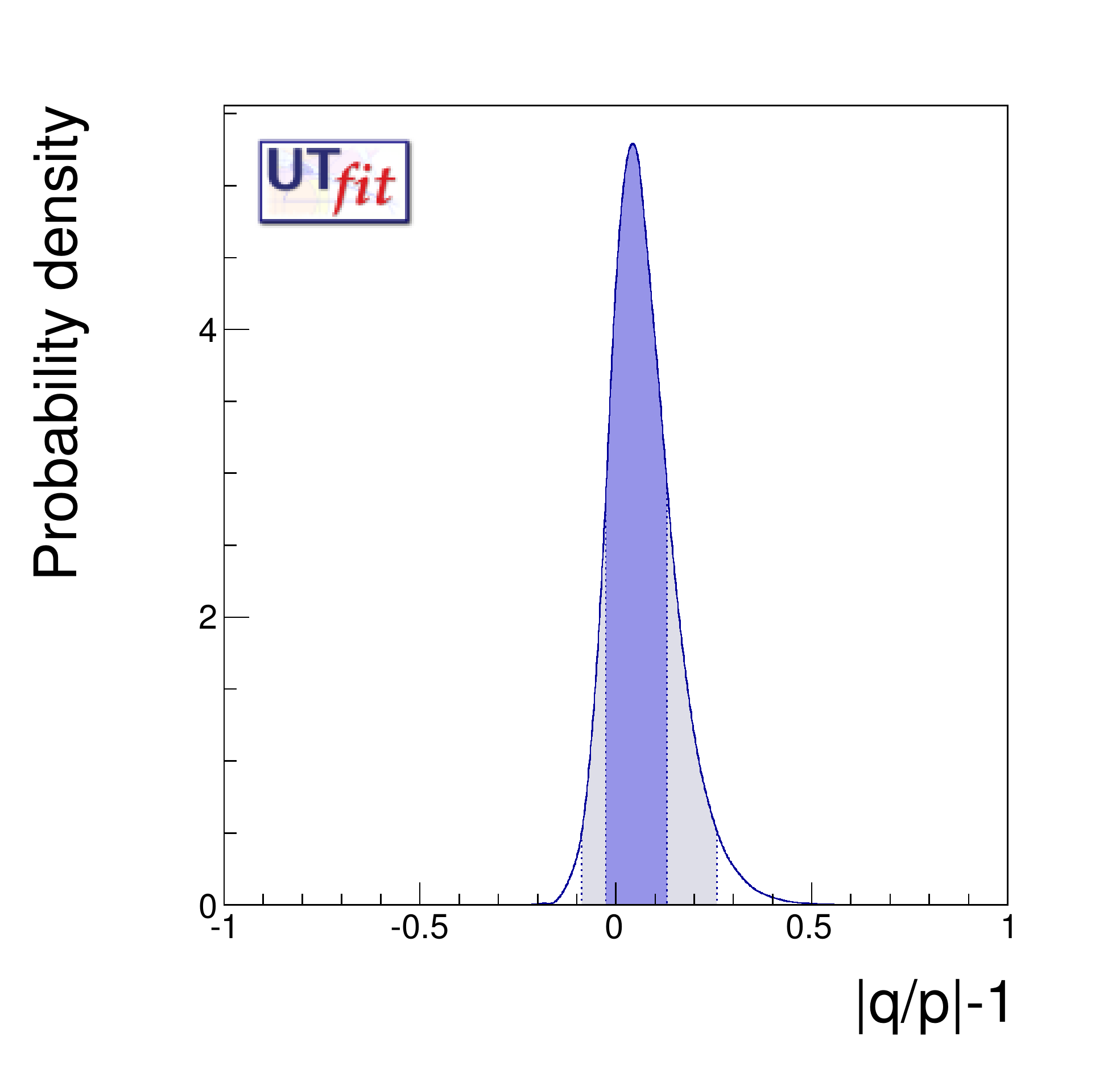}
  \includegraphics[width=.24\textwidth]{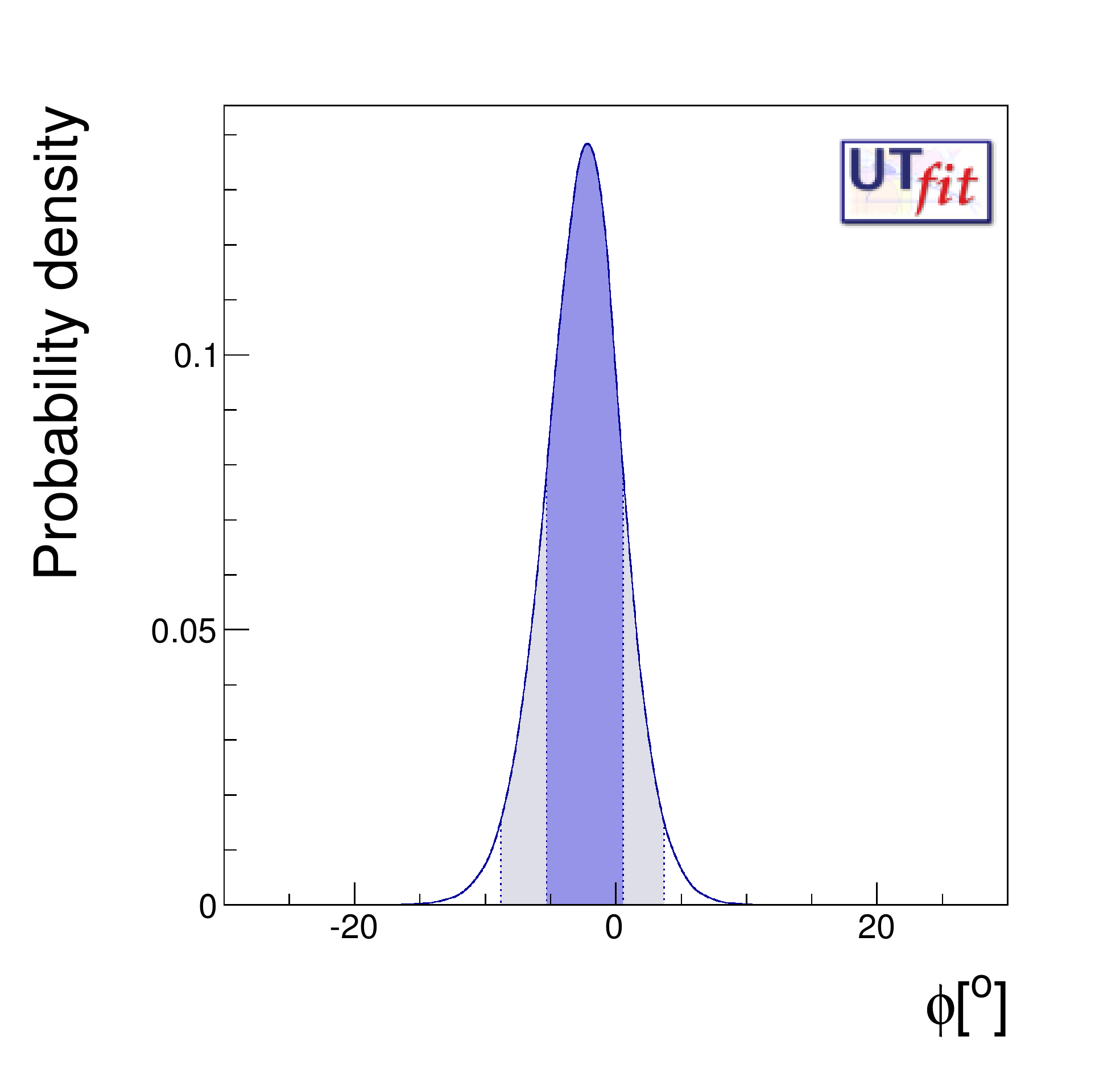}
  \caption{One-dimensional p.d.f. for the parameters $x$, $y$, $\vert
    q/p \vert -1$ and $\phi$.}
  \label{fig:ddmix_1d_2}
\end{figure}

\begin{figure}[htb]
  \centering
  \includegraphics[width=.24\textwidth]{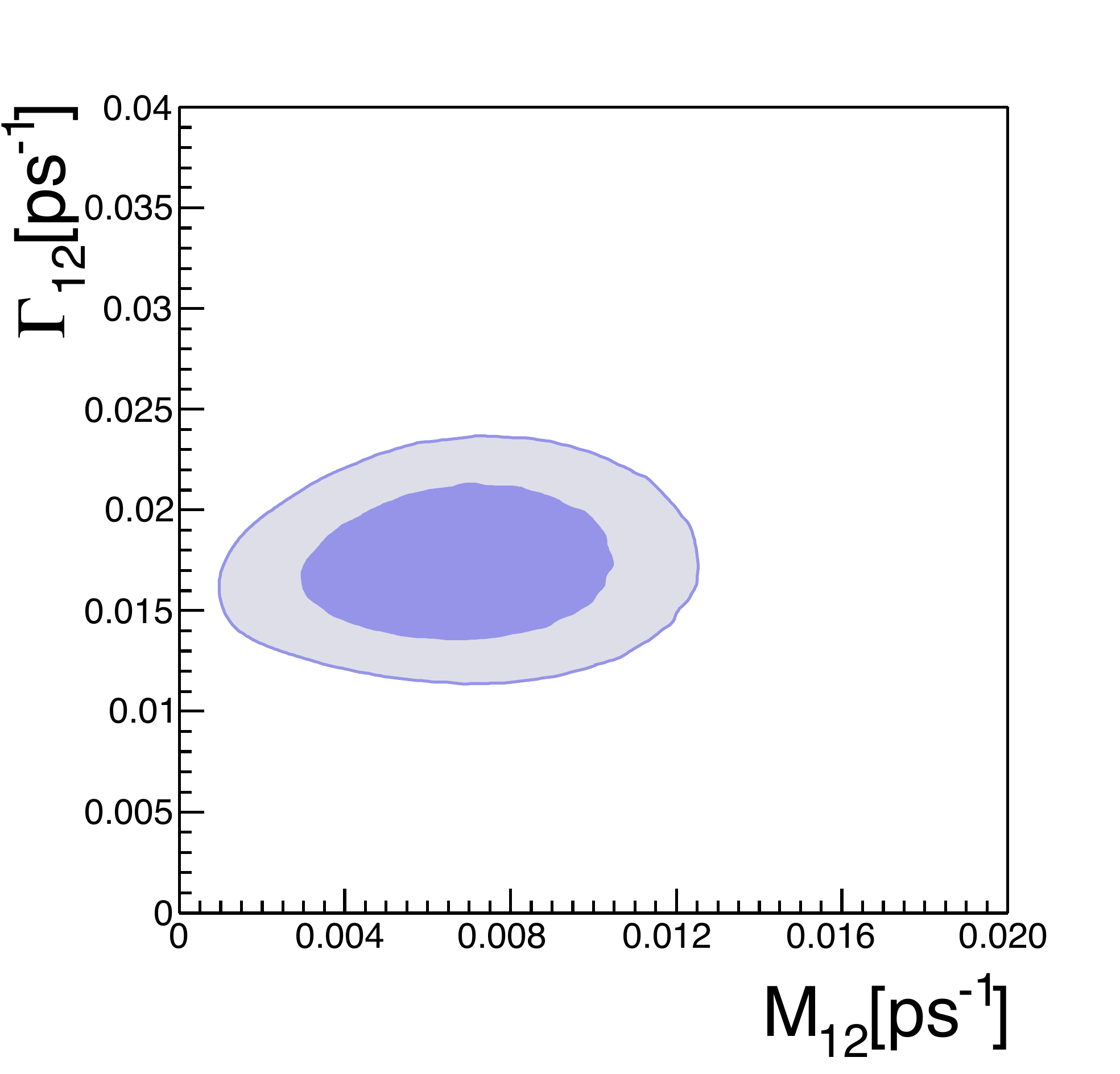}
  \includegraphics[width=.24\textwidth]{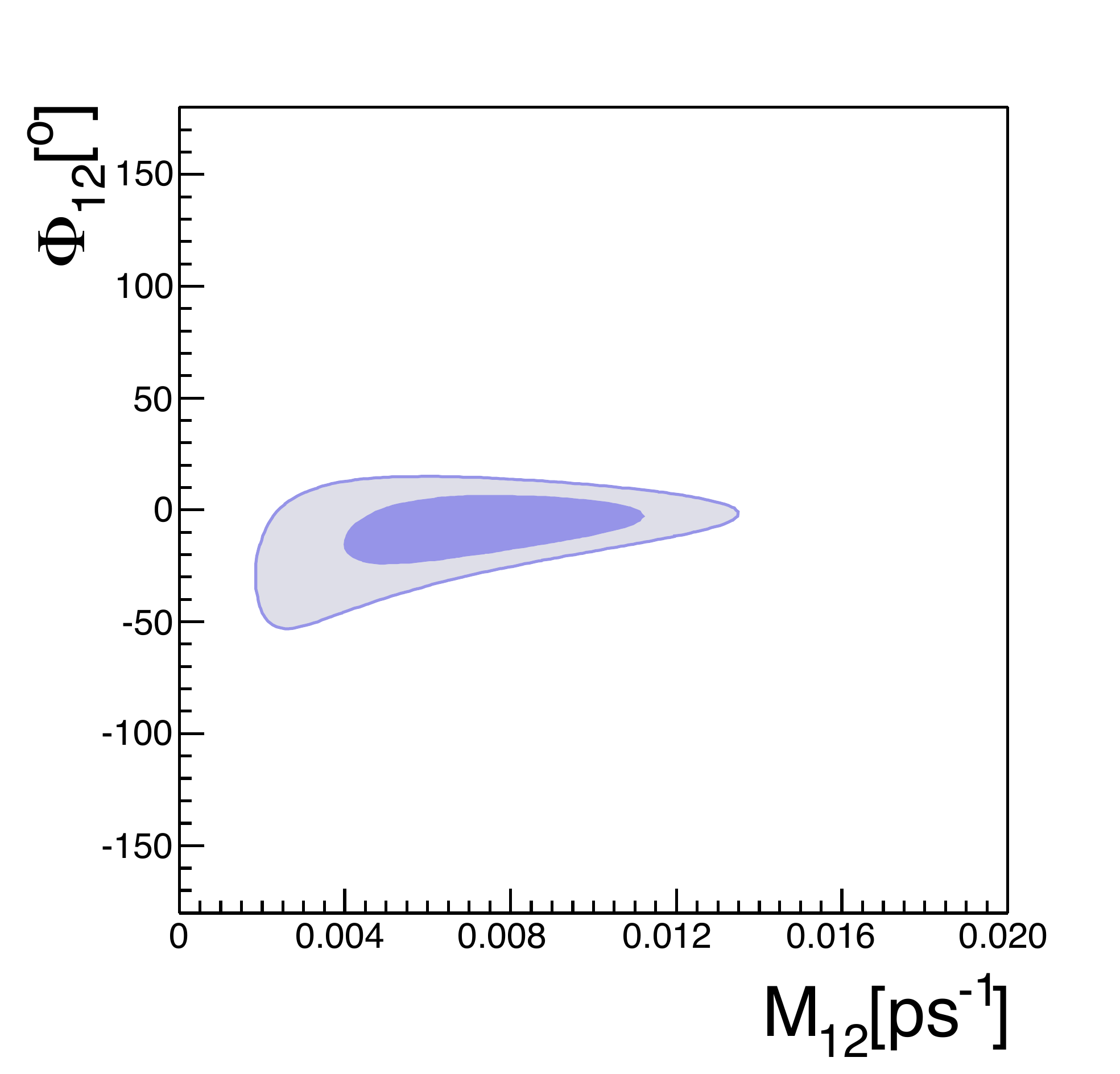}
  \includegraphics[width=.24\textwidth]{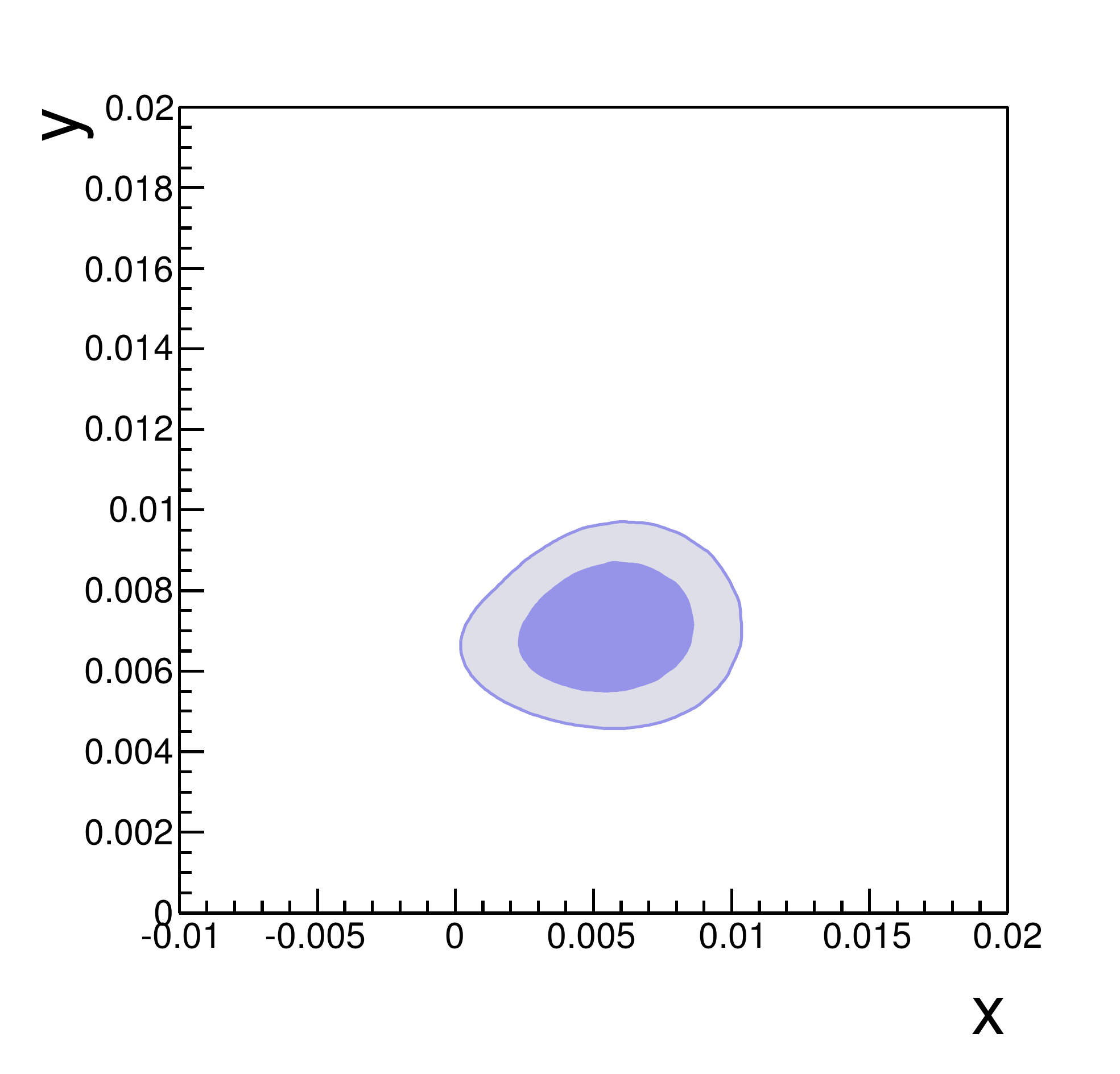}
  \includegraphics[width=.24\textwidth]{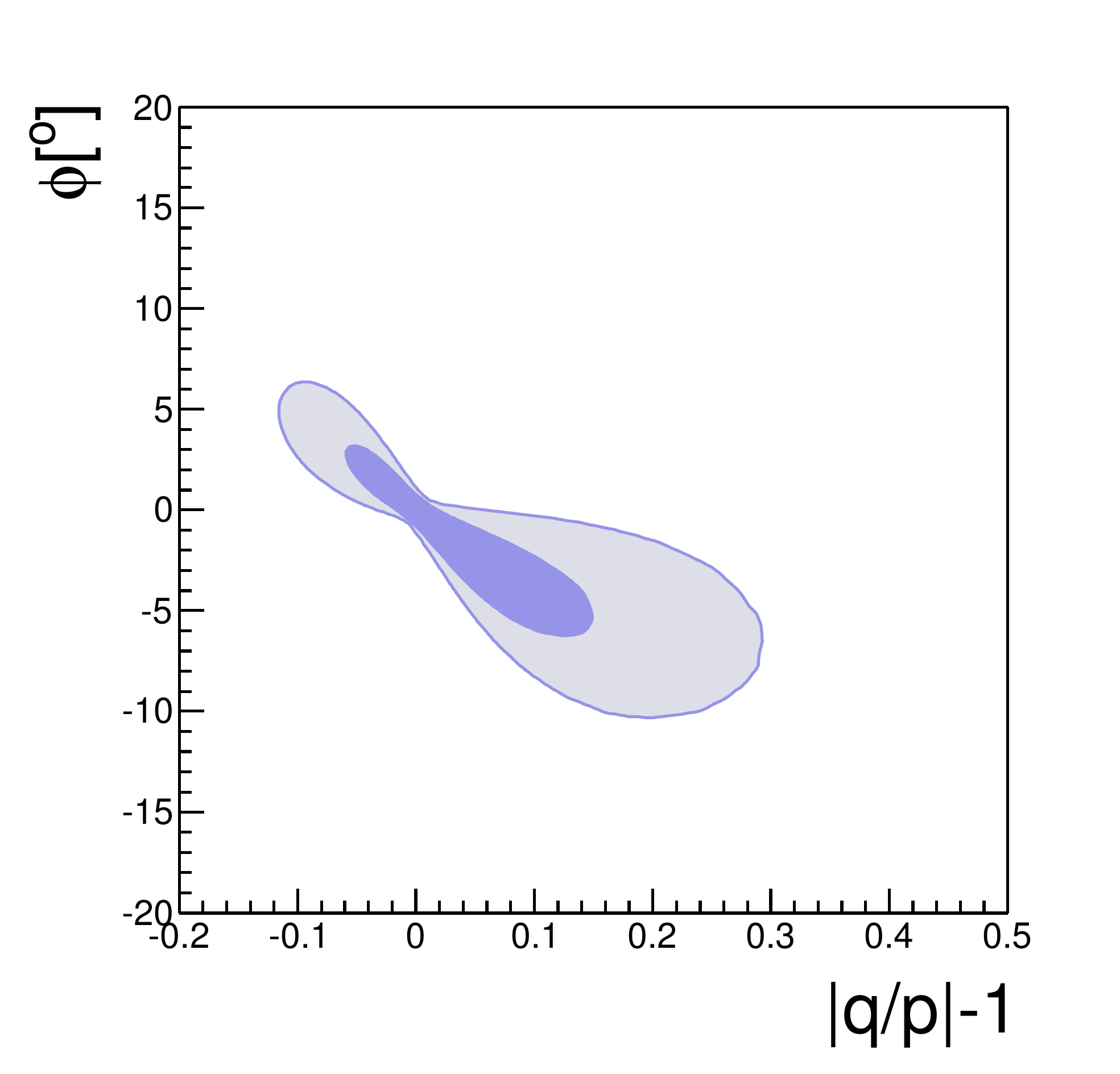}
  
  \caption{Two-dimensional p.d.f. for $\vert \Gamma_{12} \vert$ vs
    $\vert M_{12} \vert$ (top left), $\Phi_{12}$ vs
    $\vert M_{12} \vert$ (top right), $y$ vs $x$ (bottom left) and
    $\phi$ vs $\vert q/p \vert -1$ (bottom right).}
  \label{fig:ddmix_2d}
\end{figure}

The results of the fit are reported in Table \ref{tab:ddmix_res}. The
corresponding p.d.f are shown in Figs. \ref{fig:ddmix_1d} and
\ref{fig:ddmix_1d_2}. Some two-dimensional correlations are displayed
in Fig. \ref{fig:ddmix_2d}. 

A direct comparison with the HFAG results \cite{[{}][{ and online
    updates at
    \url{http://www.slac.stanford.edu/xorg/hfag/}}]1010.1589} is not
straightforward, as our fit does not fall into any of the HFAG
categories (no CPV, no direct CPV, direct CPV), since we allow for
direct CP violation only in singly Cabibbo suppressed decays. However,
our fit results should be close to the ``no direct CPV'' HFAG
fit. Indeed, we find compatible results within errors. We notice,
however, that HFAG performs a fit with four independent parameters
($x$, $y$, $\phi$ and $\vert q/p \vert$), while only three of these
parameters are independent, as can be seen from
eq.~(\ref{eq:xyandco}). In particular, $\phi$ should vanish for $\vert
q/p\vert=1$. This feature can be seen in Fig.~\ref{fig:ddmix_2d} (up
to the smoothing of the p.d.f) but not in the equivalent plot from
HFAG, which displays completely different 2-dimensional contours. We
can but recommend that in the future HFAG takes the 
relation $\phi = \mathrm{arg}(y + i \delta x)$ always into
account. 

The results in Table \ref{tab:ddmix_res} can be used to constrain NP
contributions to $D - \bar D$ mixing and decays. 

M.C. is associated to the Dipartimento di Fisica, Universit\`a di Roma
Tre. E.F. and L.S. are associated to the Dipartimento di Fisica,
Universit\`a di Roma ``La Sapienza''. We acknowledge partial support
from ERC Ideas Starting Grant n.~279972 ``NPFlavour'' and ERC Ideas
Advanced Grant n.~267985 ``DaMeSyFla''. We thank B.~Golob and
A.~Schwartz for clarifications about the HFAG averages.

\bibliography{hepbiblio}

\end{document}